# A Comprehensive Eco-Driving Strategy for Connected and Autonomous Vehicles (CAVs) with Microscopic Traffic Simulation Testing Evaluation

Ozgenur Kavas-Torris[a,*], Levent Guvenc[a]

[a] Automated Driving Lab, Department of Mechanical and Aerospace Engineering, The Ohio State University, Columbus, OH, 43212, USA.

*Corresponding author.; E-mail address: Ozgenur Kavas-Torris (kavastorris.1@osu.edu),

**ABSTRACT**

In this paper, a comprehensive Eco-Driving strategy for CAVs is presented. In this setup, multiple driving modes calculate speed profiles ideal for their own set of constraints simultaneously to save fuel as much as possible, while a High Level (HL) controller ensures smooth transitions between the driving modes for Eco-Driving. This Eco-Driving deterministic controller for an ego CAV was equipped with Vehicle-to-Infrastructure (V2I) and Vehicle-to-Vehicle (V2V) algorithms. Simulation results are used to show that the HL controller ensures significant fuel economy improvement as compared to baseline driving modes with no collisions between the ego CAV and traffic vehicles while the driving mode of the ego CAV was set correctly under changing constraints.

*Keywords:* **eco-driving, ecological cooperative adaptive cruise control, velocity trajectory, dynamic programming, traffic simulation.**

# 1. Introduction

Fuel economy enhancement in road vehicles is a problem that researchers around the world have been working to improve for decades. Eco-Driving is a term used to describe the energy efficient use of road vehicles. Some researchers have focused on improving the powertrain efficiency to improve fuel economy in vehicles (Saboohi and Farzaneh, 2009) - (Emekli and Aksun Güvenç, 2017), whereas others have worked on utilizing CAV technologies for the same purpose (Vahidi and Sciarretta, 2018), (Typaldos et al., 2020), (Fredette and Ozguner, 2018). Longitudinal autonomy and connectivity have also been utilized to achieve fuel savings for individual and platooning vehicles (Yang et al., 2020). Robust control and model regulation were also used for vehicle control (Aksun-Guvenc et al., 2003), (Aksun-Güvenç and Guvenc, 2002), (Güvenç and Srinivasan, 1994) . Parameter space with robustness was also utilized as another method for vehicle control (Güvenç et al., 2017), (Aksun-Guvenc and Guvenc, 2001), (Aksun-Guvenc and Guvenc, 2002), (Emirler et al., 2016), (Orun et al., 2009), (Demirel and Güvenç, 2010), (Oncu et al., 2007).

Developments in Vehicle-to-Infrastructure (V2I) communication technology have enhanced the capabilities of CAVs. Energy consumption of vehicles is one of the areas in the automotive industry, where connectivity can be utilized for fuel efficient driving. One way that the CAVs benefit from the V2I technology is using the infrastructure information that they receive to reduce their energy usage, whether it is fuel consumption or battery power. CAVs can use the infrastructure information, such as the locations of traffic lights and STOP signs, as well as the Signal Phase and Timing (SPaT) information from traffic lights to execute vehicle speed modification with longitudinal vehicle control in order to consume less fuel.

Altan *et al.* (Altan et al., 2017) quantified the performance of a V2I application called GlidePath with experimental testing to see how it improved fuel economy at one signalized intersection. Cantas *et al.* (Cantas et al., 2019c) and Kavas-Torris *et al.* (Kavas-Torris et al., 2020) studied and quantified the fuel consumption reduction performance of the V2I application Pass-at-Green (PaG) through extensive Monte Carlo simulations and Hardware-in-the-loop (HIL) tests for a single vehicle with no traffic around it. Kavas-Torris *et al.* (Kavas-Torris et al., 2020) also analyzed the performance of PaG employing microscopic traffic simulations with different levels of traffic. Sun *et al.* (Sun et al., 2018) used a data-driven approach for CAVs to generate an optimal speed through signalized intersections that showed 40% less fuel consumption. Asadi and Vahidi (Asadi and Vahidi, 2011) used short range radar and traffic signal information from an upcoming traffic light to generate a fuel optimal speed profile for CAVs that reduced idle time and fuel consumption. Li *et al.* (2018) utilized traffic light information to reduce idling at traffic lights and improve fuel efficiency. Li *et al.* (2015) focused on Eco-Departure from a signalized intersection for CAVs with internal combustion engines, and utilized V2I technology for Eco-Driving.

Drivers interact with other drivers during daily driving activities and are bound by the speed of a slower preceding vehicle that they are following. To consider Eco-Driving of a CAV in traffic, the preceding vehicle information also has to be taken into account by control algorithms in the ego CAV, such as the lead vehicle position and speed. Vehicles can also communicate with each other through V2V to get acceleration information and use it for fuel economy, emissions and safety benefits.

Cruise Control (CC) systems aim to keep the vehicle speed constant to aid the drivers on roadways and are particularly helpful for freeway driving (Bjornberg, 1994). CC design has usually employed classical control methods while approaches like fuzzy logic control (Shaout and Jarrah, 1997) have also been used. They help in safety and are useful as Driver Assist Systems (DAS), however, CC models do not adjust the ego vehicle speed with respect to outside input, such as preceding vehicle position and speed.

Adaptive Cruise Control (ACC) has been widely used for saving fuel and improving safety for vehicles (Xiao and Gao, 2010) (Pan et al., 2021). ACC is an important part of the Advanced Driver Assistance Systems (ADAS) and SAE Level 2 automated vehicles are equipped with ACC for car following scenarios ("J3016B," n.d.). An ego vehicle equipped with a classical ACC uses cameras and radars to detect and track the preceding vehicle, and actuators to control the ego vehicle speed (Pan et al., 2021). Kural and Aksun-Guvenc designed an ACC model using Model Predictive Control (MPC) (Kural and Aksun Güvenç, 2010). By reducing the unnecessary accelerations and decelerations as much as possible in the ego vehicle, ACC systems help to improve performance and indirectly save fuel. However, V2V technology is not utilized in ACC systems.

Cooperative Adaptive Cruise Control (CACC) enables V2V to be used for car following scenarios (Wang et al., 2018). In CACC, the ego vehicle receives information about the preceding vehicle from the preceding vehicle itself via V2V communication. Hu *et al.* (Hu et al., 2017) developed an optimal look-ahead control framework that improved fuel economy in car following situations using V2V technology. Cantas *et al.* (Cantas et al., 2019b) implemented a CACC system for an ego vehicle that used the preceding vehicle acceleration information through V2V communication. Kianfar *et al.* (Kianfar et al., 2012) designed a CACC architecture that is capable of driving within a vehicle platoon while minimizing inter-vehicular spacing within allowed ranges, attenuating shock waves and ensuring safety. Rasool *et al.* (Rasool et al., 2019) used Pontryagin's Minimum Principle (PMP) to improve fuel efficiency during car following with CACC. Guvenc *et al.* (Güvenç et al., 2012) designed and tested a CACC system for the Grand Cooperative Driving Challenge (GCDC). Naus *et al.* (Naus et al., 2010) used the frequency-domain approach to design and experimentally validate a string-stable CACC system.

Ecological Cooperative Adaptive Cruise Control (Eco-CACC) is an improvement over the CACC system and aims to improve fuel efficiency using road information while utilizing CACC in car following scenarios or vehicle platoons. Zhai *et al.* (Zhai et al., 2020) designed an Eco-CACC model for a heterogeneous platoon of heavy-duty vehicles with time delay between the platoon agents. Yang *et al.* (Yang et al., 2017) modeled an Eco-CACC algorithm to compute the fuel-optimum vehicle trajectory through a signalized intersection that also handles queue effects. Almannaa *et al.* (Almannaa et al., 2019) designed an Eco-CACC model to reduce fuel consumption and achieve travel time savings around signalized intersections, and also tested the system through field implementation.

In this study, a comprehensive Eco-Driving strategy was developed for a CAV equipped with V2I and V2V algorithms. This study shows relative fuel savings each component provides to CAVs, how each component can be improved and what constitutes the largest effect on fuel savings. It has been shown that the complete Eco-Driving architecture presented in this paper is applicable to be used in real life in actual vehicles. The main contributions of this paper can be summarized as follows:

1) V2I and V2V algorithms were developed to control the longitudinal motion of a CAV for Eco-Driving.

2) The Higher Level (HL) controller was also tested in a traffic simulator with realistic traffic flow. The traffic vehicles were controlled by the traffic simulator, and had default car following models, which enabled them to change lanes when they were behind slower vehicles. Thus, the traffic vehicles created dynamically changing constraints on the HL controller. It was observed that the HL controller ensured that no collisions were observed between the ego CAV and traffic vehicles, and the driving mode of the ego CAV was set correctly under changing constraints.

3) The High Level (HL) controller designed for the comprehensive Eco-Driving of a CAV enabled fuel savings.

The rest of the paper is organized as follows. Section 2 describes the comprehensive Eco-Driving strategy for CAVs that was developed in this work. Section 3 details the fuel optimal Eco-Cruise driving mode and the deterministic High Level controller. The microscopic traffic simulation environment is introduced in Section 4. Section 5 discusses the simulation results and comparative analysis based on various performance measures, followed by conclusions summarized in Section 6.

**2. Complete Eco-Driving Strategy for a Connected and Automated Vehicle (CAV)**

The schematic in Figure 1 displays a complete picture for the comprehensive Eco-Driving strategy for CAVs proposed in this paper. Firstly, the CAV needs to have a speed profile, which is called Eco-Cruise, that is route dependent and fuel optimal. This Eco-Cruise speed profile would assume normal operating conditions, meaning it would assume no surrounding traffic and infrastructure around the CAV. Additionally, speed limit of the route, as well as safe acceleration and deceleration limits for ride comfort, need to be enforced as constraints during the calculation of this fuel optimal speed profile. This speed profile takes the route elevation into account, as well as the constraints of the vehicle, and can be calculated offline by using Dynamic Programming (DP). The Eco-Cruise mode shown in Figure 1 is the default driving mode, meaning that when the ego CAV does not interact with other vehicles or is not in the vicinity of traffic signs, Eco-Cruise is active to consume as little fuel as possible.

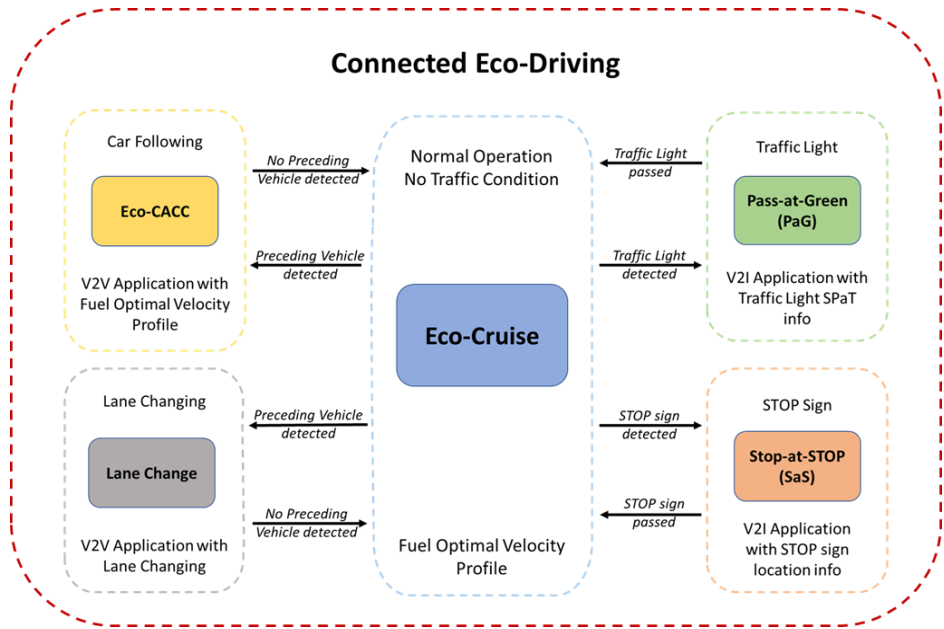

Figure 1. Comprehensive Eco-Driving architecture of CAVs

CAVs interact with roadway infrastructure, such as traffic lights and STOP signs. For the Eco-Driving of a CAV, when there is an upcoming traffic light and the traffic light Signal Phase and Timing (SPaT) information is broadcast from a Roadside Unit (RSU), then the ego CAV goes into Pass-at-Green (PaG) mode (green colored block in Figure 1). In this mode, the ego CAV picks up the SPaT information from the upcoming traffic light and determines whether it is able to pass the light by accelerating to a higher speed, keeping its speed constant or decelerating to a lower speed. If one of these three states is possible, then the PaG calculates a smooth speed profile for the ego CAV to follow, so that fuel economy and ride comfort are maximized. For the state where the vehicle is not able to pass, then the PaG calculates a smooth Eco-Approach to the traffic light, so that the vehicle decelerates smoothly and spends as little time as possible while idling during the red light. Once the light turns green, then the PaG calculates a smooth Eco-Departure speed profile from the traffic light.

For the Eco-Driving of a CAV, the ego CAV also interacts with STOP signs on roadways. STOP signs are usually not equipped with any type of V2I equipment, therefore another tool needs to be used to get STOP sign location information. In this architecture, the ego CAV is equipped with eHorizon (Autoliv Inc., 2020), an electronic horizon that has a detailed map in it. Once the ego CAV gets close to the STOP sign location, it goes into the Eco-Stop mode (red colored block in Figure 1). In the Eco-Stop mode, using the STOP sign location information, an Eco-Approach speed profile is calculated that enables the ego CAV to decelerate smoothly in a fuel optimal manner, and come to a complete stop at the STOP sign. After the ego CAV waits for 5 seconds at the STOP sign during the simulations, the Eco-Departure is subsequently activated to get the vehicle to speed up to the speed limit. A perception sensor like a camera and image processing should be used in conjunction with the electronic horizon map in practice to be certain of the STOP sign location and presence. While a 5 second wait period is fine for the fuel economy computations in this paper, the CAV should use perception and communication sensors to assess safety of operation before proceeding to depart the STOP sign.

Other than the roadway infrastructure, CAVs also interact with other surrounding traffic agents. CAVs are equipped with perception sensors, hence, they can detect nearby objects or vehicles. For the Eco-Driving of a CAV, once the ego CAV detects a preceding vehicle, then it needs to go into the Eco Cooperative Adaptive Cruise Control (Eco-CACC) mode (light orange colored block in Figure 1). This mode uses V2V communication, so that the ego CAV gets preceding vehicle information and uses that information to follow the preceding vehicle in a fuel efficient manner.

When the preceding vehicle movement is too erratic or the preceding vehicle is moving too slowly, then the ego CAV goes into Lane Change mode (gray colored block in Figure 1). In the Lane Change mode, the ego CAV gets surrounding vehicle information, such as vehicle speed and acceleration, as well as vehicle position. Then, the model determines if it is safe to change lanes and executes lane changing. The main goal of Lane Change in Eco-Driving of a CAV is to make sure the ego CAV can maintain the optimal Eco-Cruise speed to get maximum fuel savings.

**2.1 Vehicle-to-Infrastructure (V2I) Interactions of a CAV**

A vehicle travelling from a starting location to a traffic light (or STOP sign) can be seen in Figure 2. In the figure, $x_{ego}$ is the position, $v_{ego}$ is the speed and $a_{ego}$ is the acceleration of the ego vehicle, respectively. $TL_{location}$ is the traffic light location, $TL_{SPaT}$ is the signal phase and timing of the traffic light and $STOP_{location}$ is the location of the STOP sign.

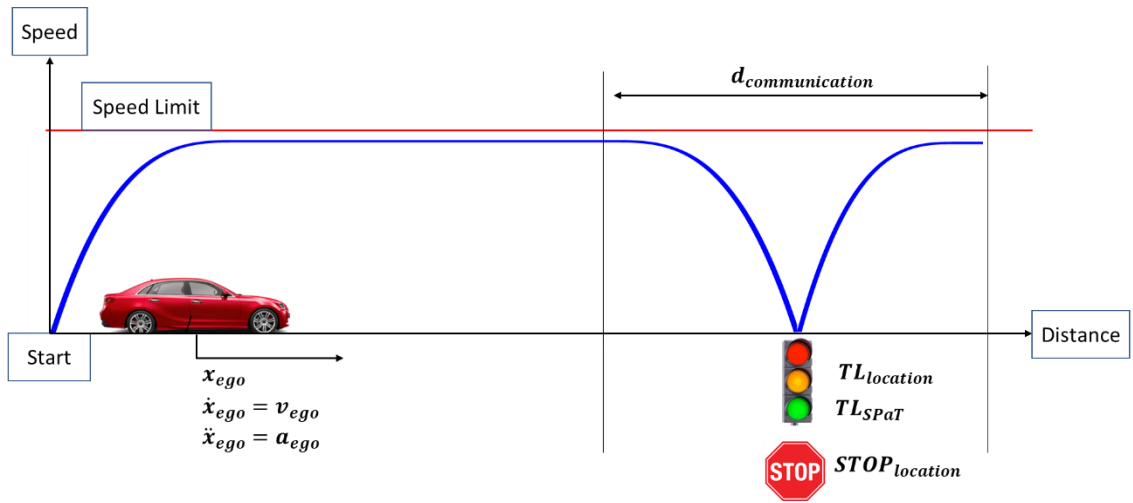

Figure 2. V2I interaction as an optimal control problem

In this paper, the aim is to design control algorithms that minimize the fuel consumption in a vehicle. One way to reduce fuel consumption is by the utilization of V2I, so that the vehicle control algorithms can get roadway infrastructure information and use it to consume less fuel. The optimal control problem can be defined with the objective function (1),

$$\underset{T_e(t), F_b(t)}{minimize}\ J(u(t)) = L_N\big(s(t_f), v(t_f)\big) + \int_0^{t_f} L_k(s(t), v(t), T_e(t), F_b(t), t) dt \qquad (1)$$

where $L_k$ is the running cost and $L_N$ is the terminal cost. The states are subject to,

$$\frac{ds(t)}{dt} = v(t) \qquad (2)$$

$$\frac{dv(t)}{dt} = K_{T_e} T_e - K_{F_b} F_b - g r_0 \cos(\alpha(t)) - \frac{1}{2m} \rho_{air} A_f C_D v(t)^2 - g\sin(\alpha(t)) \qquad (3)$$

where equation (2) expresses that the derivative of position is equal to the speed. Equation (3) shows that road load and brake force subtracted from the total powertrain force to the wheels is equal to the vehicle acceleration. There are initial and final algebraic constraints on the states of position and speed, and they are as follows,

$$s(0) = s_{initial} = 0 \qquad (4)$$

$$s(t_f) = s_{final} = s_f \qquad (5)$$

$$s_{min}(t) \leq s(t) \leq s_{max}(t) \qquad (6)$$

$$v(0) = v_{initial} = v_i = 0 \qquad (7)$$

$$v(t_f) = v_{final} = v_f = 0 \qquad (8)$$

$$v_{min}(t, s(t)) \leq v(t) \leq v_{max}(t, s(t)) \qquad (9)$$

where $s_{initial}$ (4) is the initial vehicle position, $s_f$ (5) is the final vehicle position, $s_{min}$ is the minimum position for the vehicle and $s_{max}$ is the maximum position for the vehicle (6). Additionally, $v_i$ (7) is the initial vehicle speed, $v_f$ (8) is the final vehicle speed and $v_{min}$ (9) is the minimum allowable speed for the vehicle. The speed limit of the roadway is enforced as the $v_{max}$ (9) constraint, which is the maximum allowable speed of the vehicle. There are also algebraic constraints on inputs engine torque $T_e$ (10) and brake force $F_b$ (11),

$$T_{e,min}(v(t),t) \leq T_e(t) \leq T_{e,max}(v(t),t) \tag{10}$$

$$0 \leq F_b(t) \leq F_{b,max}(v(t),t) \tag{11}$$

The optimal control problem posed here was solved using Dynamic Programming and was called the Eco-Cruise speed, which will be covered in section 3.

**2.2 Vehicle-to-Vehicle (V2V) Interactions of a CAV**

An ego CAV following a lead connected vehicle can be seen in Figure 3. $x_{ego}$ and $x_{lead}$ are the positions of the ego and lead vehicles, respectively. $\dot{x}_{ego}$ and $\dot{x}_{lead}$ are the speeds of the ego and lead vehicle, respectively. $\ddot{x}_{ego}$ and $\ddot{x}_{lead}$ are the accelerations of the ego and lead vehicle, respectively. It should be noted that the sinusoidal looking perturbation in the speed profile of Figure 3 is for illustration purposes only and represents a perturbation (not necessarily sinusoidal) that the ego vehicle does not want to follow.

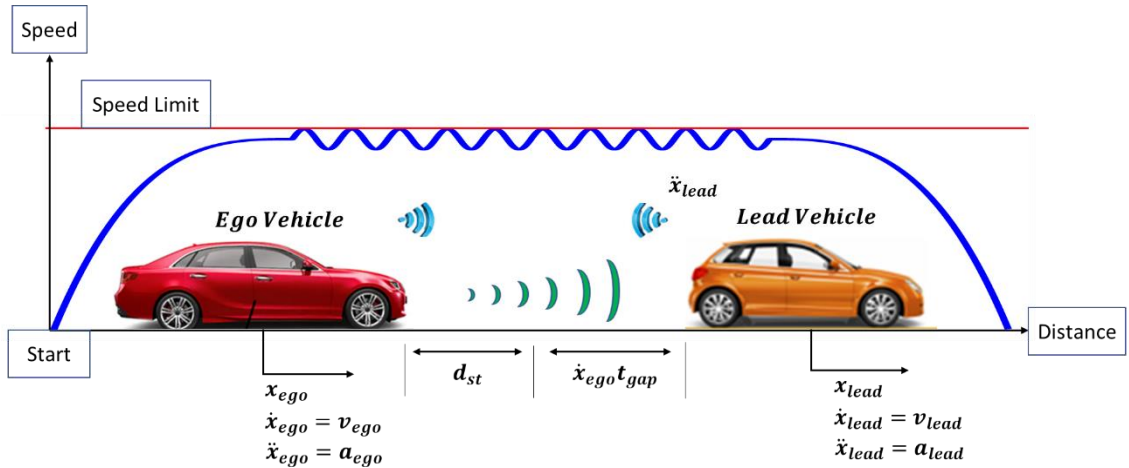

Figure 3. Car following of a CAV as an optimal control problem

Fuel consumption in CAVs can be reduced by the utilization of V2V, so that the vehicle control algorithms can get lead vehicle information and use it to consume less fuel. For this optimal control problem, similar to the case for V2I interactions of a CAV, equations (1) to (11) can be defined.

Other algebraic constraints also need to be enforced, so that a collision between the ego and the lead vehicle does not occur. These constraints are as follows

$$x_{actual} = x_{lead} - x_{ego}, x_{actual} > 0 \qquad (12)$$

where $x_{actual}$ (12) is the actual distance between the lead and the ego vehicle, and it needs to be always larger than zero to ensure that the vehicles do not collide.

## 3. Eco-Driving Modes and the High Level Controller

In this section, the algorithms that were developed for the Eco-Driving of a CAV are explored further.

### 3.1. Fuel Optimization with Eco-Cruise

Dynamic Programming is a well-known solution that is used to find optimal benchmark solutions to various optimal control problems. A generic Dynamic Programming (DP) Matlab function dpm [9] was used in the calculation of the fuel optimal Eco-Cruise speed profile for a conventional vehicle. In this paper, the problem was to minimize the road load acting on the vehicle (Figure 4), so that the fuel consumed by the vehicle would also be minimized.

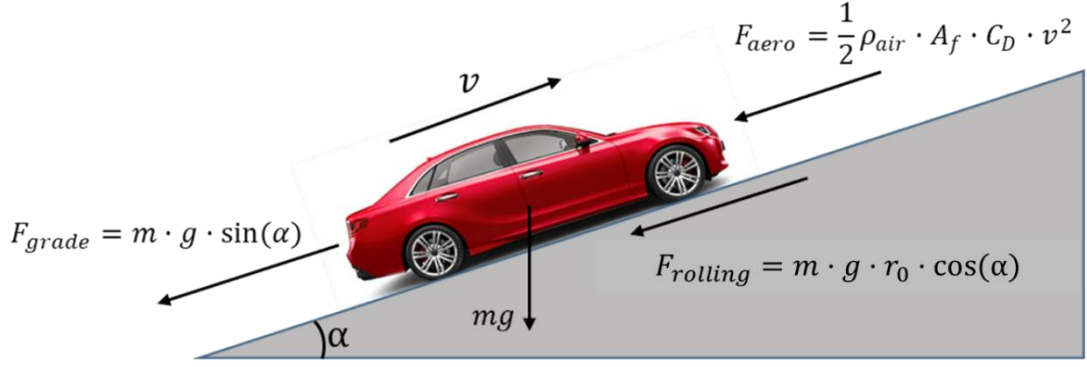

Figure 4. Road forces acting on a vehicle

$$Road\ Load = F_{rolling} + F_{aero} + F_{grade} \qquad (13)$$

$$Road\ Load = mgr_0 \cos(\alpha(s)) + \left(\frac{1}{2}\rho_{air}A_f C_D v^2\right) + mg\sin(\alpha(s)) \qquad (14)$$

The road load (Figure 4) equation given in (13) has 3 parts. The first part is the rolling resistance $F_{rolling}$, and depends on vehicle speed, tire properties and road conditions. The second term $F_{aero}$ is the aerodynamic drag term and is dependent on vehicle speed and frontal cross-section area of the vehicle. The last term is $F_{grade}$ the road grade term and is dependent on the road grade and vehicle mass. In Equation (14), $m$ is the vehicle mass with the rotating inertia factor, $r_0$ is a parameter of the rolling resistance equation, $\alpha$ is the road grade, $\rho_{air}$ is the density of air, $A_f$ is the front cross-sectional area, $C_D$ is the drag coefficient, $v$ is vehicle speed.

The power that needs to be provided from the engine in a vehicle to beat road load and enable acceleration can be expressed as follows,

$$P = F_x v = \left(m_e \frac{dv}{dt} + \frac{1}{2}\rho_{air} \cdot A_f \cdot C_D \cdot v^2 + m \cdot g \cdot r_0 \cdot \cos(\alpha) + m \cdot g \cdot \sin(\alpha)\right)v \qquad (15)$$

where $P$ is the power, $F_x$ is the force required at the tires and $m_e \frac{dv}{dt}$ is the force required to accelerate. The rest of the terms in Equation (15) come from the road load acting on the vehicle, which was given in Equation (14). Using the power expression given in Equation (15), the fuel rate that is consumed by the vehicle when it is travelling can be expressed as follows,

$$\dot{m}_f = \frac{P/\eta_t + P_{accessories}}{\eta_e} \qquad (16)$$

where $\dot{m}_f$ is the fuel rate, $P_{accessories}$ is the power required to keep the accessories running, $\eta_t$ is the transmission efficiency and $\eta_e$ is the engine efficiency. This expression for the fuel rate given in Equation (16) can be used as the cost function that needs to be minimized for this analysis. Further details for this optimal control formulation can be found in (Kavas-Torris et al., 2021).

In this paper, the fuel optimal DP solution presented here was used for different driving modes. Firstly, the driving mode called Eco-Cruise, where the fuel optimal speed profile is calculated offline using road information, was found using DP. Additionally, the Eco-Stop mode, where the ego vehicle approaches a STOP sign fuel economically, also utilized DP. Finally, the Eco-Departure mode, where the ego vehicle departs from a traffic light or STOP sign, also used DP. These solutions were all distance-based solutions, as presented earlier.

In the DP solution, the whole horizon is divided into segments and the solution space is partitioned into nodes. Starting from the desired end point, a cost is assigned to move from the current node to each previous neighboring node in a backward propagation approach. Then, the feasibility constraint of going from one node to the next is enforced. As seen in Figure 5, the sequence of feasible propagation nodes of minimum cost is then formed as the minimum cost solution (Kavas-Torris et al., 2021).

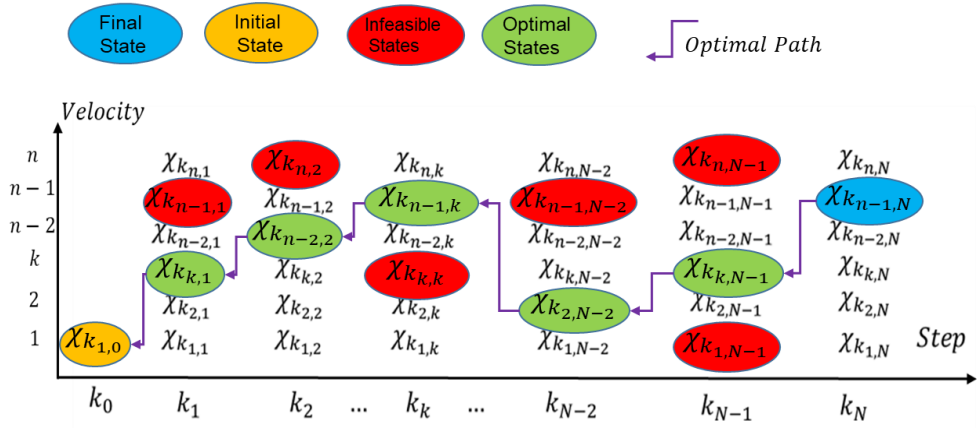

Figure 5. DP algorithm table for the optimal solution (Kavas-Torris *et al.*, 2021)

## 3.2. High Level (HL) Controller for V2I with No Traffic

The High Level (HL) controller handles how the ego CAV behaves when it is travelling on a roadway with no other vehicle around it and is implemented as a state flow chart. The aim is to determine when the CAV has to switch between the different driving modes of the Eco-Driving of a CAV architecture presented in Figure 1. This controller ensures seamless transition from one driving mode to the next.

Depending on deterministic conditions, such as the current upcoming traffic light state and duration, the distance between the infrastructure elements (traffic lights and STOP signs) and the ego vehicle, as well as the instantaneous vehicle speed, the controller is tasked to make a decision to switch between driving modes. The flow chart for the deterministic control algorithm for the fuel economic Eco-Driving of a single CAV with no traffic can be seen in Figure 6.

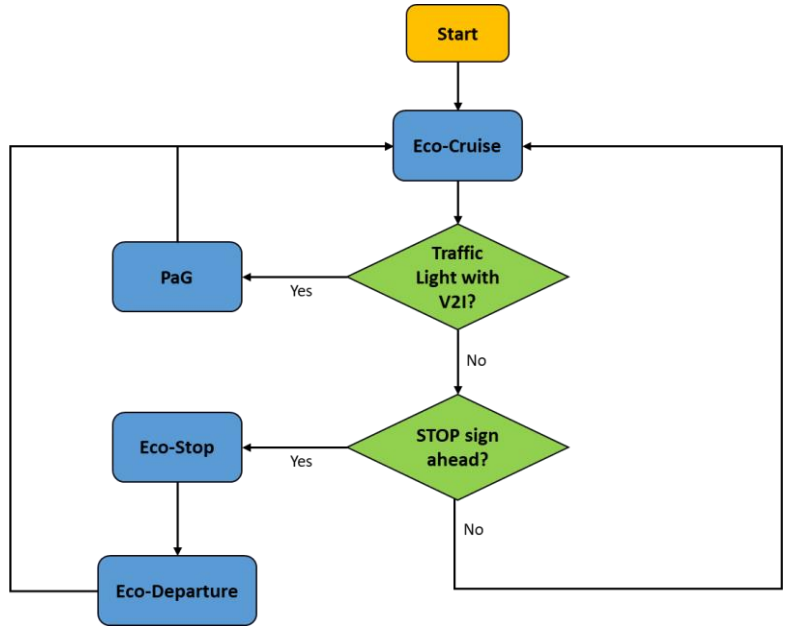

Figure 6. Flowchart for the HL controller with no traffic

As seen in Figure 6, the ego CAV aims to maintain its speed as close to the Eco-Cruise speed as possible. The Eco-Cruise speed is the fuel economic speed profile that is route dependent and is calculated offline prior to the trip. If there is an upcoming traffic light, then Pass-at-Green (PaG) V2I algorithm takes over control of the ego vehicle. If there is a STOP sign, then Eco-Stop mode is activated to make the vehicle stop smoothly at the sign. After stopping at the STOP sign for a few seconds, then Eco-Departure takes over and makes the ego vehicle accelerate smoothly. The HL controller makes sure the correct driving mode is active and mode transitions are smooth to save as much fuel as possible.

The Pass-at-Green (PaG) is a V2I application that uses roadway infrastructure information to eliminate or decrease idling at red lights to decrease the fuel consumption for the ego vehicle (Altan et al., 2017) (Cantas et al., 2019c) (Kavas-Torris et al., 2020). PaG operates under deterministic control by using the input which are the distance to the upcoming traffic light, Signal Phase and Timing (SPaT) information received from the upcoming traffic light, instantaneous actual vehicle speed, maximum acceleration and maximum deceleration limits and jerk limit for ride comfort. Using these inputs, PaG calculates a smooth and fuel economic speed profile, so that the vehicle can pass the upcoming traffic light.

**3.2. High Level (HL) Controller for V2V with Traffic**

This High Level (HL) controller aims to make transitions between driving modes correctly and smoothly, so that the ego vehicle speed does not jump abruptly when the driving mode changes. The flowchart for the controller is seen in Figure 7, where the Eco-Cruise speed is the fuel economic and road dependent speed profile for the ego vehicle to follow to consume less fuel. When there is a preceding vehicle with no V2V communication, then the Adaptive Cruise Control (ACC) model is activated, and the ego CAV safely follows the lead vehicle. If the preceding vehicle is equipped with V2V and is not an erratic driver, then the Cooperative Adaptive Cruise Control (CACC) takes over and follows the lead vehicle smoothly while keeping a safe distance between vehicles to prevent collision. If the preceding vehicle with V2V is an erratic driver, then Ecological Cooperative Adaptive Cruise Control (Eco-CACC) takes over control to follow the erratic leader without responding to its high frequency accelerations in order to maintain fuel savings and safety. If the leader is erratic and lane changing is possible for the ego vehicle, then the ego vehicle changes its lane and maintains Eco-Cruise speed.

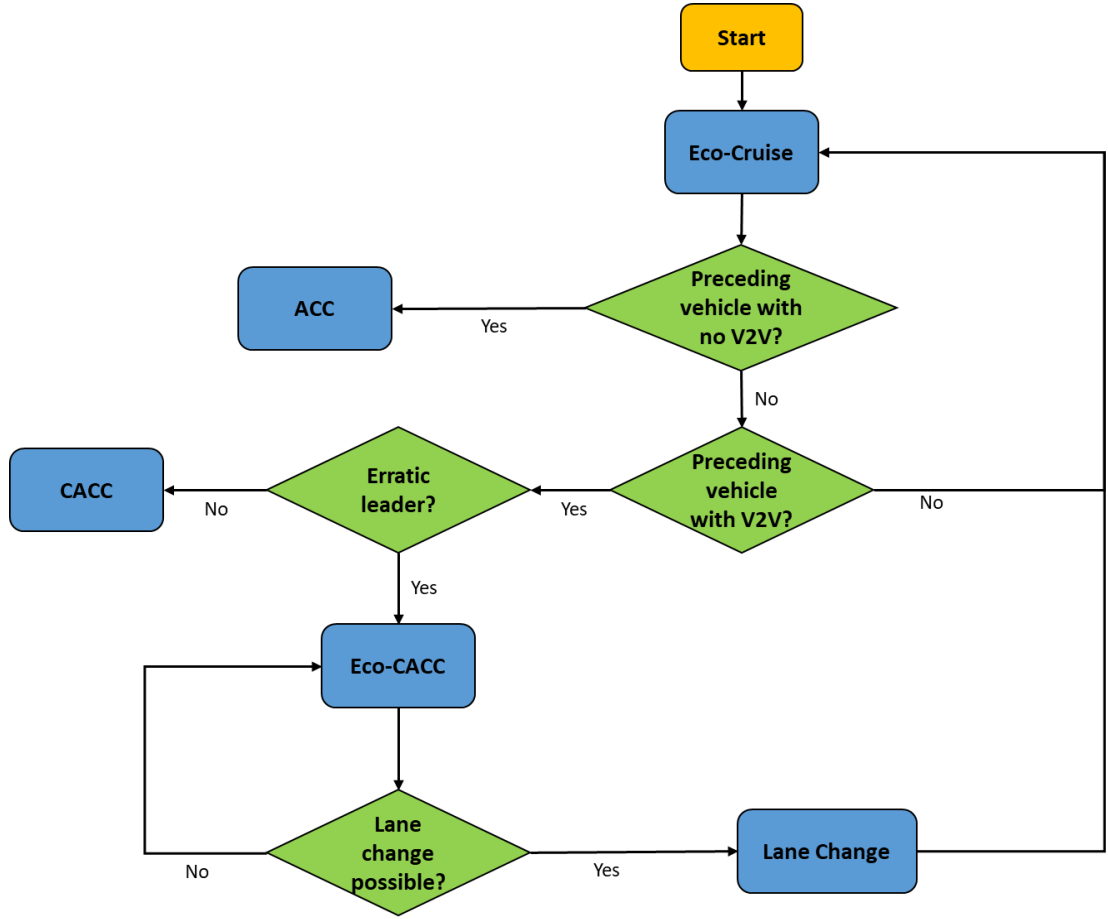

Figure 7. Flowchart for the HL controller with traffic

Driving modes shown in Figure 7 have different controllers and when they all run simultaneously during testing, the recommended vehicle speeds from each driving mode are usually different. If driving modes were to switch immediately with no transition, then recommended speeds would not be continuous and cause the actual ego vehicle speed to jump abruptly. To overcome this problem, a Transition State is added to smoothly transition between driving modes. The algebraic equation for the Transition State to smoothly increase the vehicle speed is as follows,

$$v_{trans} = v_{trig} + v_{chg,acc} \qquad (17)$$

$$v_{chg,acc} = (v_{lim} - v_{trig})\left(\left(\frac{t_{act} - t_{trig}}{4(v_{lim} - v_{trig})} - 1\right)^3 + 1\right) \quad (18)$$

where $v_{trans}$ (17) is the recommended transition speed for the vehicle, $v_{trig}$ is the vehicle speed when driving mode transition started, $v_{chg,acc}$ is the speed change needed for the ego vehicle to travel at the higher speed limit. In Equation (18), $v_{lim}$ is the actual speed limit of the road, $t_{act}$ is the actual simulation time and $t_{trig}$ is the time instant when driving mode transition starts. The 3$^{rd}$ order power equation that comprises of the variables seen in Equation (18) ensures that the recommended speed is smooth when driving modes are switched and the ego CAV accelerates.

When the Eco-Cruise speed is smaller than the instantaneous vehicle speed, then the following Equation (19) ensures the vehicle decelerates slowly. In Equation (19), $v_{chg,dec}$ is the speed change needed for the ego vehicle to travel at the lower speed limit. In Equation (20), $v_{lim,low}$ is the user set lower speed limit, $t_{act}$ is the actual simulation time and $t_{trig}$ is the time instant when driving mode transition starts. A 3$^{rd}$ order power equation that comprises of the variables seen in Equation (20) ensures that the recommended speed is smooth when driving modes are switched and the ego CAV decelerates. When the Eco-Cruise speed catches up to the vehicle speed, then the recommended speed for the vehicle to follow switches back to the Eco-Cruise speed.

$$v_{trans} = v_{trig} - v_{chg,dec} \quad (19)$$

$$v_{chg,dcc} = (v_{lim,low} - v_{trig})\left(\left(\frac{t_{act} - t_{trig}}{4(v_{lim,low} - v_{trig})} - 1\right)^3 + 1\right) \quad (20)$$

## 4. Microscopic Traffic Simulation Environment

A simulation environment was set up between Simulink and Vissim commercial traffic simulator to run co-simulations using the COM interface capability of Vissim (Kavas-Torris et al., 2020), (Cantas et al., 2019a). During the co-simulations, realistic traffic information was being sent from Vissim to Simulink. The ego vehicle with a midsized vehicle powertrain was being controlled by the High Level (HL) Controller in Simulink. The HL controller determined which action to take and which driving mode to activate in response to the realistic traffic and infrastructure information received from the traffic simulator.

The simulation environment designed in Vissim has one STOP sign, five traffic lights and the total length of the route called Arlington Route is 6,873 m. The speed limit, traffic sign locations and route-dependent fuel economic DP solution for the Eco-Cruise driving mode for Route Arlington can be seen below in Figure 8.

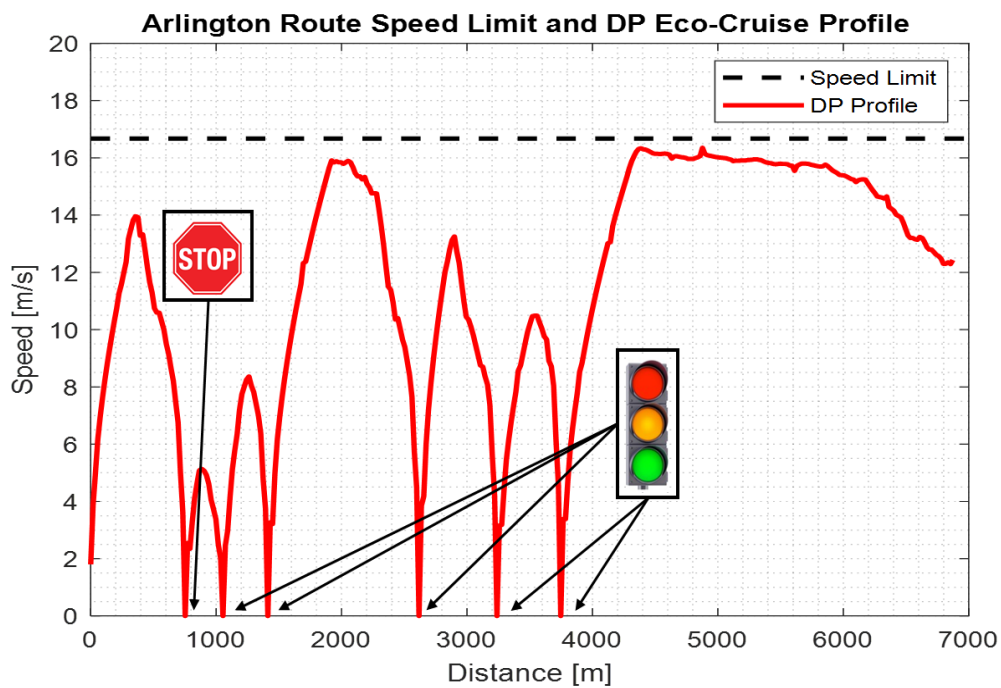

Figure 8. Characteristics of Arlington Route

The HL controller was designed as a state flow diagram in Simulink and the flowchart for the HL controller decision making process can be seen in Figure 9. The default mode is the Eco-Cruise mode, where the pre-calculated fuel economic DP profile is the desired speed profile for the vehicle. The Eco-Cruise speed profile also makes sure the ego vehicle drives in a fuel economic manner around STOP signs. When there is a lead vehicle in close proximity to the ego vehicle, then Car Following mode are activated to safely and closely follow the preceding vehicle. When there is a traffic light ahead, the mode is switched to PaG. After the ego vehicle passes the traffic light, then depending on the instantaneous speed of the vehicle, the transition modes are activated (Speed Up or Speed Down).

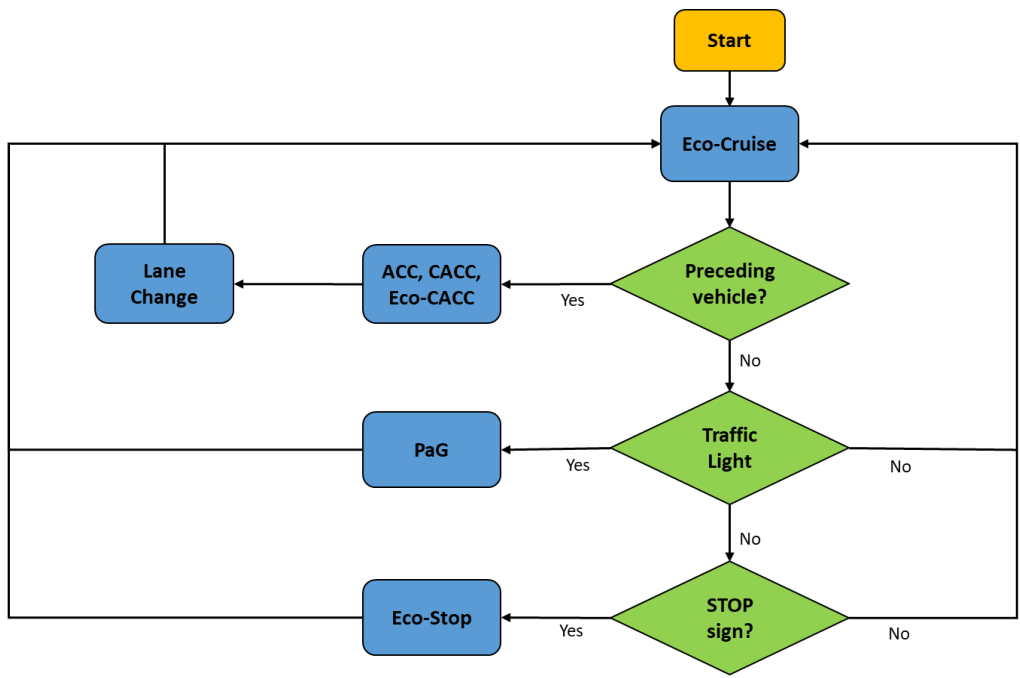

Figure 9. HL controller flowchart for traffic simulations

The pink ego vehicle approaching a traffic light at an intersection with other traffic vehicles around it during the traffic simulation can be seen in Figure 10. During the simulation, the ego vehicle was controlled by the HL controller to save fuel by smoothly approaching traffic lights and STOP signs. At the same time, whenever there was a vehicle in front of the ego vehicle, and the distance between the ego and lead vehicles was less than 50 m, then ACC, CACC or Eco-CACC were activated to prevent collision between the vehicles during car following.

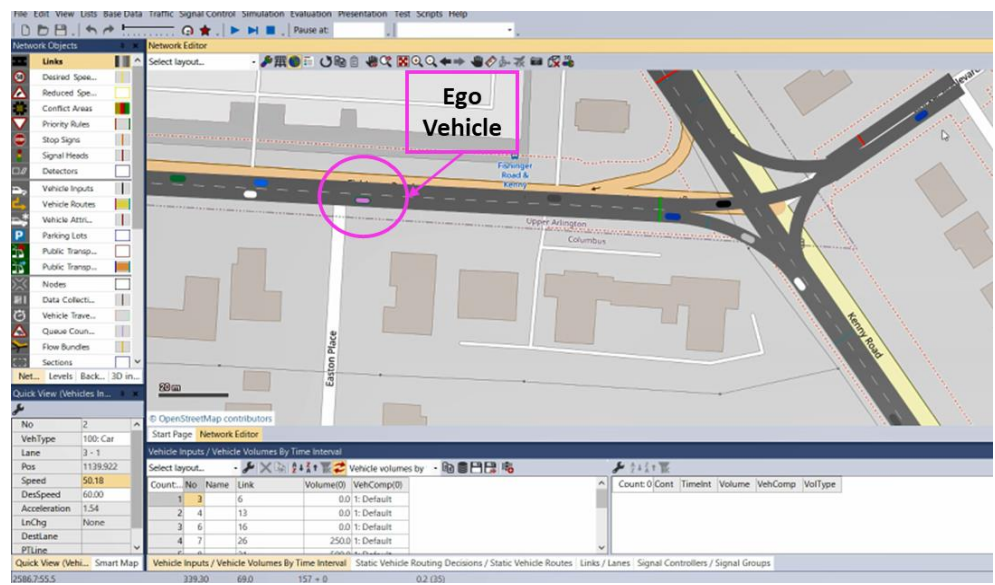

Figure 10. Ego vehicle approaching an intersection in the traffic simulation

The traffic vehicle compositions were the same at each simulation. Additionally, the traffic simulator spawned vehicles at a common start time for each simulation, meaning that the vehicle with a specific ID entered the roadway at the same timestamp across all simulation cases. This unity ensures that the simulation results can be compared with each other, since the traffic vehicles that interact with the ego vehicle appear in the simulator at the same time. Additionally, the traffic light periods for each traffic light were the same across all simulations.

## 5. Results and discussion

To assess the fuel economy performance of the V2I and V2V algorithms in a traffic network, 5 different simulations were run. For the 1st case, the ego vehicle was commanded to follow the fuel economic DP profile for the Eco-Cruise mode with no other traffic vehicles around in the simulation. This first case was set as the baseline for the fuel economy comparisons that were later conducted in the analysis. For the 2nd case, the ego vehicle was commanded to follow the same Eco-Cruise speed as the 1st case while also interacting with STOP signs using Eco-Stop and traffic lights using PaG. The 3rd simulation case built on top of the second simulation case, where Eco-Cruise, Eco-Stop and PaG were all working in tandem, and there were also traffic vehicles around the ego vehicle. Whenever the ego vehicle was in the vicinity of a lead vehicle, then the ACC mode was activated. The 4th simulation case used the same V2I models, and when there was a lead vehicle ahead, the CACC mode was activated. The 5th and final simulation case used the same V2I models as the 4th case, except the car following model that was used when there was a lead vehicle in front of the ego vehicle for this case was Eco-CACC.

The speed profile for the ego vehicle when there were no other traffic vehicles around can be seen in Figure 11. The light blue line represents the ego vehicle speed when it was commanded to follow the DP offline calculated Eco-Cruise profile in Figure 8. The light red line represents the vehicle speed when the vehicle was in the vicinity of a traffic light, and the SPaT information was used to modify the speed profile for case 2.

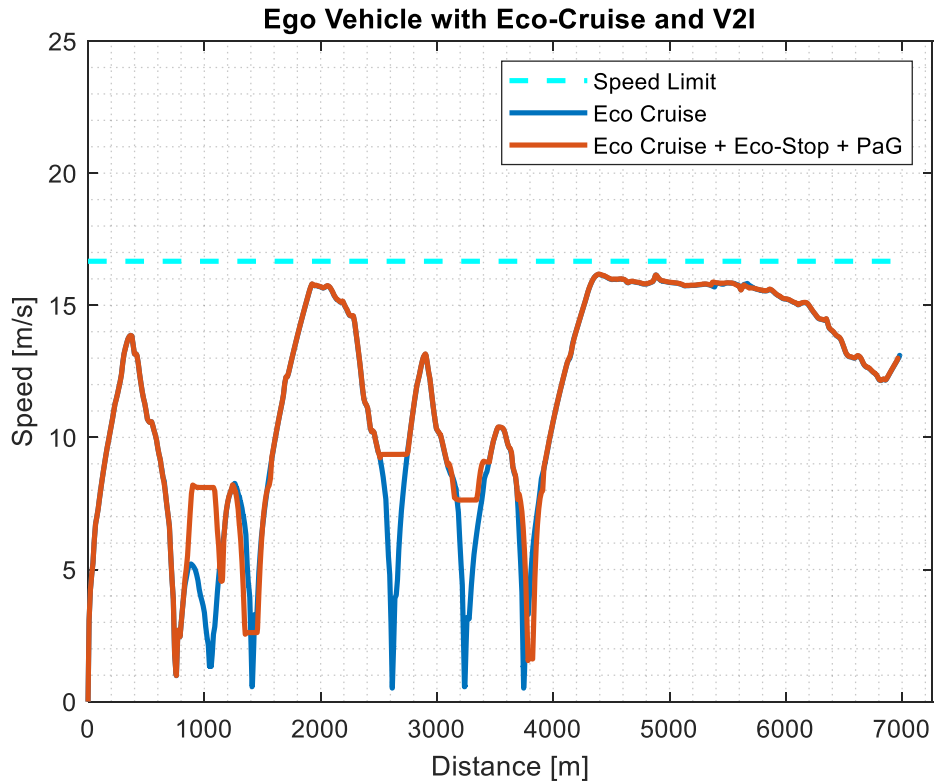

Figure 11. Ego vehicle with Eco-Cruise only and Eco-Cruise + Eco-Stop + PaG

The results for the 3rd simulation case, where there were other traffic vehicles in the traffic simulator and the ego vehicle was equipped with the V2I algorithms and ACC can be seen in Figure 12. Whenever the distance between the ego vehicle and the lead vehicle was below 50 m, the ACC took over the control to make sure no collision could occur. If the distance between the ego and the lead vehicle was larger than 50 m, then the HL controller commanded the ego vehicle to either follow the Eco-Cruise trajectory or the PaG trajectory to save fuel.

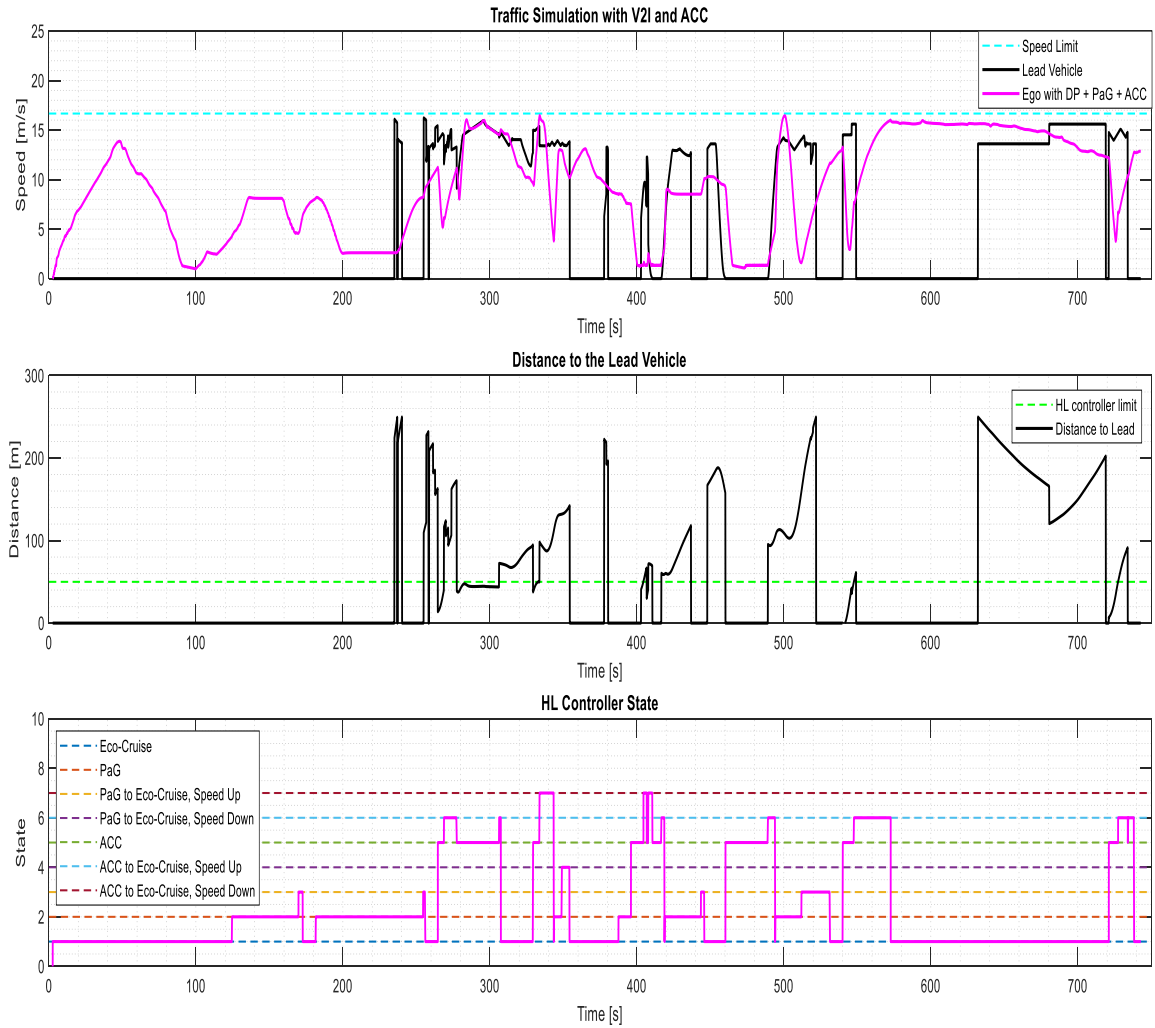

Figure 12. Traffic simulation for ego vehicle with V2I and ACC

The results for the 4th simulation case, where there were other traffic vehicles in the traffic simulator and the ego vehicle was equipped with the V2I algorithms and CACC can be seen in Figure 13. The HL controller handled having a preceding vehicle ahead of the ego vehicle the same as the ACC case.

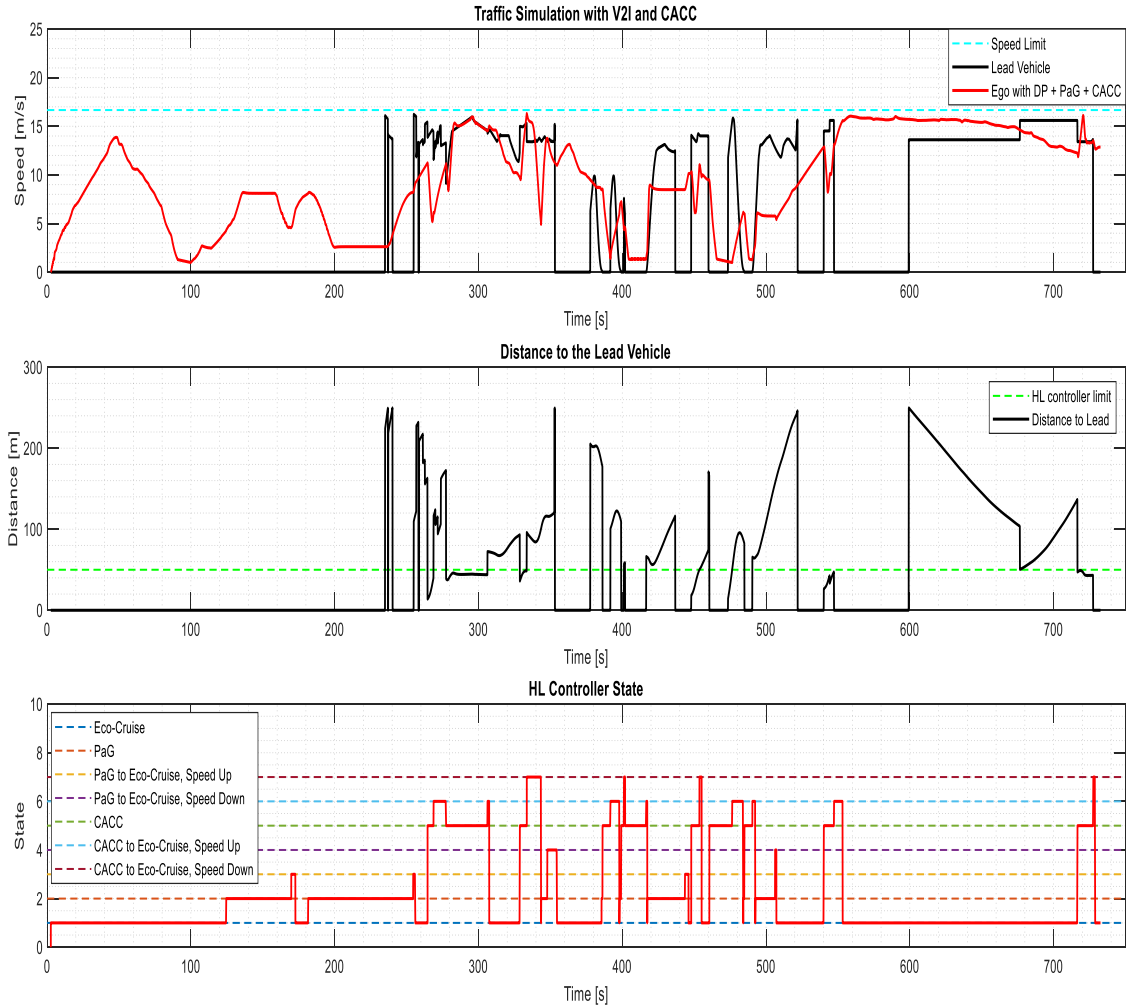

Figure 13. Traffic simulation for ego vehicle with V2I and CACC

The results for the 5th and the final simulation case, where there were other traffic vehicles in the traffic simulator and the ego vehicle was equipped with the V2I algorithms and Eco-CACC can be seen in Figure 14. Similar to the previous cases with ACC and CACC, the HL controller handled the state transitions.

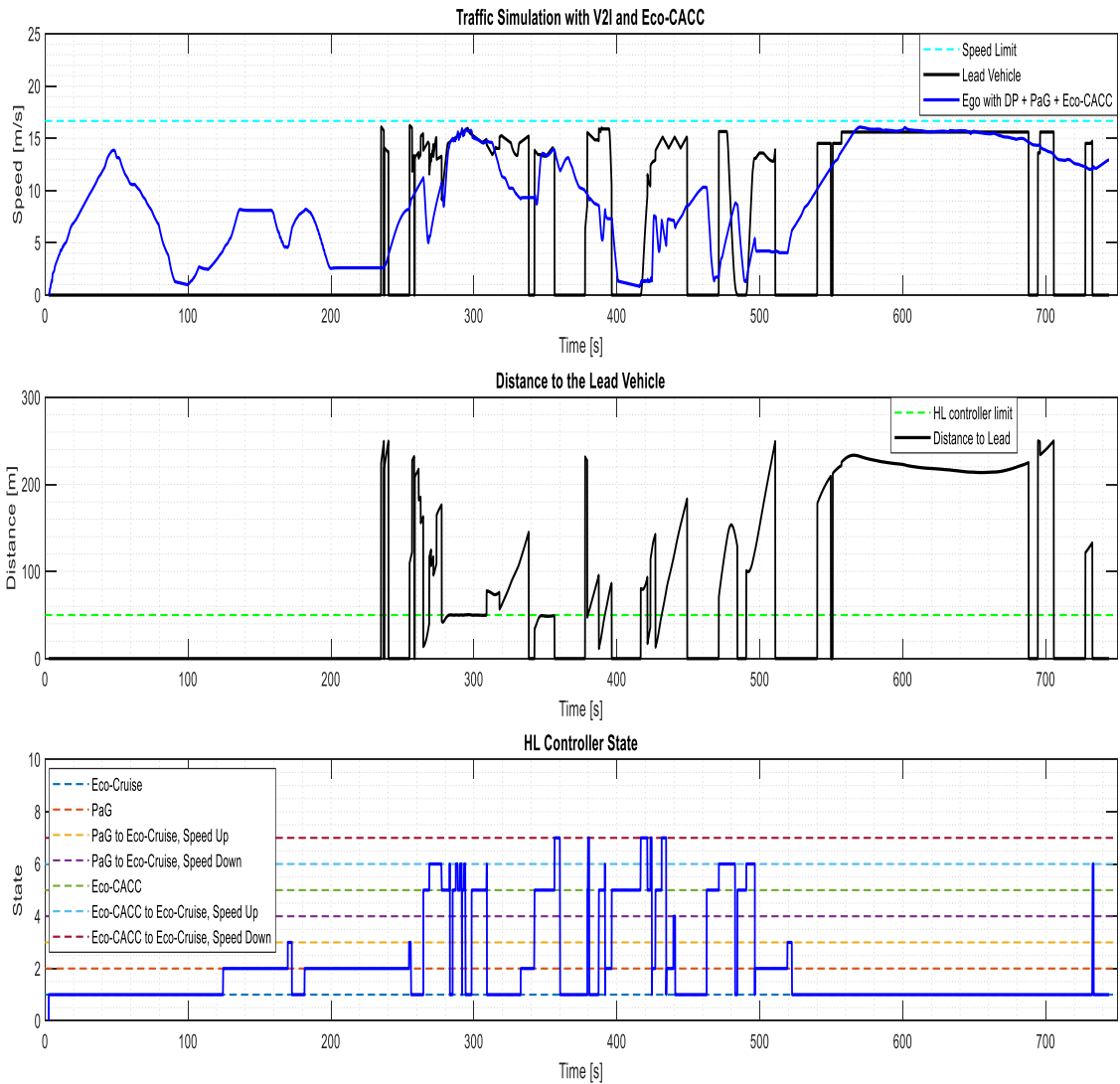

Figure 14. Traffic simulation for ego vehicle with V2I and Eco-CACC

The fuel consumed for the ego vehicle for each of the five simulation cases were recorded and the percentage fuel consumption reduction of the models were calculated with respect to the simulation case 3 (Table 1). During case 1 and case 2, there were no other vehicles around the ego vehicle to interact with using V2V, therefore, the fuel consumption for case 1 and case 2 are smaller compared to the case 3. When the ego vehicle could use V2I in case 2, the fuel consumed by the ego vehicle decreased compared to using Eco-Cruise only simulation in case 1, where the ego vehicle stops at all traffic lights and STOP signs. Traffic vehicles that constrained the motion of the ego vehicle were present for cases 3, 4 and 5. Compared to ACC for car following in case 3, using CACC in case 4 resulted in a 1.51% fuel economy improvement. Moreover, using the Eco-CACC in case 5 was even more beneficial in reducing the fuel consumed by the ego vehicle. The fuel consumption decreased by 6.41% using Eco-CACC in case 5 compared to using ACC in case 3.

Table 1. Results for the fuel economy of the ego vehicle in traffic network

| Simulation Case Number | Simulation Scenario Name | Total Fuel Consumption [g] | % Fuel Consumption Reduction wrt Case #3 |
|---|---|---|---|
| 1 | Eco-Cruise only (no traffic, vehicle stops at all traffic lights) | 395.85 | 12.85% |
| 2 | Eco-Cruise with Eco-Stop and PaG (no traffic) | 382.17 | 15.86% |
| 3 | Eco-Cruise with Eco-Stop and PaG and ACC | 454.20 | - |
| 4 | Eco-Cruise with Eco-Stop and PaG and CACC | 447.37 | 1.51% |
| 5 | Eco-Cruise with Eco-Stop and PaG and Eco-CACC | 425.12 | 6.41% |

## 6. Conclusions

In this paper, a comprehensive Eco-Driving strategy with V2I and V2V algorithms was tested in a realistic microscopic traffic simulation environment. When PaG was active and used traffic infrastructure information, 3.46% less fuel was consumed compared to only using Eco-Cruise speed profile. For the simulation cases that required car following, it was shown that using CACC and Eco-CACC with V2V were more beneficial than using only ACC. The ego vehicle consumed 1.51% and 6.41% less fuel as compared to ACC only for car following when CACC and Eco-CACC were used, respectively. Moreover, it was seen that Eco-CACC, which was modeled with a filter to attenuate the acceleration of the lead vehicle, consumed less fuel than CACC, which used the lead vehicle acceleration without filtering it.

## Acknowledgments

This work is supported by the Automated Driving Laboratory (ADL) at The Ohio State University (OSU).